\documentclass[aps,pre,groupedaddress,amsmath,amssymb]{revtex4}
    
\usepackage{amsmath}
\usepackage{graphicx}

\begin{document}
\title{Non-exponential relaxation for anomalous diffusion}

\author{M. H. Vainstein}
\affiliation{International Center of Condensed Matter Physics and Institute of Physics, University of Bras\'{\i}lia - CP 04513 70919-970 Bras\'{\i}lia-DF, Brazil}

\author{I. V. L. Costa}
\affiliation{International Center of Condensed Matter Physics and Institute of Physics, University of Bras\'{\i}lia - CP 04513 70919-970 Bras\'{\i}lia-DF, Brazil}

\author{R. Morgado}
\affiliation{International Center of Condensed Matter Physics and Institute of Physics, University of Bras\'{\i}lia - CP 04513 70919-970 Bras\'{\i}lia-DF, Brazil}

\author{F. A. Oliveira}
\email[Corresponding author: ]{fao@fis.unb.br}
\affiliation{International Center of Condensed Matter Physics and Institute of Physics, University of Bras\'{\i}lia - CP 04513 70919-970 Bras\'{\i}lia-DF, Brazil}

\begin{abstract}
We study the relaxation process in normal and anomalous diffusion regimes 
for systems described by a generalized Langevin equation (GLE). We demonstrate the existence of a very general correlation function which describes
the relaxation phenomena. Such function is even; therefore, it cannot be an exponential or a stretched exponential. However,
for a proper choice of the parameters, those functions can be reproduced within certain intervals with good precision. We also show the
passage from the non-Markovian to the Markovian behaviour in the normal diffusion regime. For times longer than the relaxation time,
the correlation function for anomalous diffusion becomes a power law for broadband noise. 
\end{abstract}

\maketitle

\section{Introduction\label{sec.introduction}}

The research on the striking universality properties of slow relaxation
dynamics has been driving great efforts in the last decades. A large and
growing literature can be found in which non-exponential behaviour has been observed for correlation functions: supercooled colloidal systems~\cite{Rubi04}, glasses and granular material~\cite{Santamaria-Holek04,Vainstein03a}, liquid crystal
polymers~\cite{Santos00,Benmouna01}, hydrated proteins~\cite{Peyrard01},
growth~\cite{Colaiori01}, plasmas~\cite{Ferreira91}  and disordered vortex lattices in
superconductors~\cite{Bouchaud91} are just a few examples. Those systems present features which are similar to those found in systems with anomalous diffusion.  The attempt to obtain response functions which are able to explain such relaxation processes is a subject more than a hundred years old. Rudolph Kohlrausch used stretched exponentials $R(t) \sim \exp[-(t/\tau)^{\beta}]$ with $0<\beta<1$ to describe charge relaxation in a Leyden gas~\cite{Kohlrausch54}. Later on, his son, Friedrich Kohlrausch~\cite{Kohlrausch63} observed two distinct universalities: the stretched exponential with $0<\beta <2$, and the power law behaviour. The former behaviour is now known as the Kohlrausch-Williams-Watts (KWW) stretched exponential. 

There are now two main methods to describe those relaxations: a fractional derivatives approach~\cite{Chaves98,Metzler00,Metzler04}, and a recurrence method applied to the Mori equation~\cite{Balucani03}. However, new methods appear every day and we should mention here the attempt of Chamberlin~\cite{Chamberlin96} to explain this universal behaviour, the diffusive method of Perez-Madrid~\cite{Perez-Madrid04}, and the new method recently proposed by Schwartz and Edwards~\cite{Schwartz02} for non-linear fields.

In this work, we present an alternative way to obtain the correlation function and a result which is general as long as the diffusive processes can be described by a generalized Langevin equation (GLE). We derive a correlation function which is an even function of time. The asymptotic behaviour is similar to those obtained using fractional derivatives. However, for short times, our method can lead to richer behaviours.
The paper is organized as follows: first, we give the outline of the problem of a system governed by a GLE and show the connection to diffusion; we then continue by defining in a clear manner the noise, memory and correlation functions. We construct the memory function from the fluctuation-dissipation theorem using a coloured noise, from a given density of states.  After that, we discuss normal and anomalous diffusions. Finally, we analyze the difference between Markovian and non-Markovian behaviours.

\section{Generalized Langevin equations and diffusion}
We shall start by writing the GLE for an operator $A$
\begin{equation}
\label{gle}
\frac{d A(t)}{d t}=-\int _{0}^{t}\Pi (t-t')A(t')\,d t'+F(t),
\end{equation}
where $F(t)$ is a stochastic noise subject to the conditions $\langle F(t) \rangle =0$, $\langle F(t)A(0) \rangle =0 $ and
\begin{equation}
\label{fdt}
C_F(t-t')=\langle F(t)F(t') \rangle = \langle A^2 \rangle_{eq} \Pi(t-t').
\end{equation}
 Here, the angular brackets denote an ensemble average. Some correlation functions depend on two times, being consequently non-stationary~\cite{Rubi04,Santamaria-Holek04,Vainstein03a}.
Equation~(\ref{fdt}) is the Kubo fluctuation-dissipation theorem (FDT)~\cite{Kubo66,Kubo91}. The FDT can be violated in many slow relaxation processes~\cite{Costa03,Vainstein06}. 
Before continuing, it is important to stress the connection between the generalized Langevin equation for a given operator $A$ and diffusion. To do this, we define the variable 
\begin{equation}
x(t)=\int_0^t A(t')\,d t'.
\end{equation} 
Now, we can study the asymptotic behaviour of  its second moment $\lim_{t\rightarrow \infty} \langle x^2(t) \rangle \sim t^{\alpha}$ to characterize the type of diffusion presented by the system: for $\alpha = 1$, we have normal diffusion, and for $\alpha<1 (>1)$, subdiffusion (superdiffusion).

The memory kernel $\Pi (t)$ indicates that the motion is non-Markovian; when $\Pi(t)=2\gamma \delta(t)$, eq.~(\ref{gle}) becomes the usual Langevin equation.
Our main interest is to discuss the behaviour of the correlation function\[
R(t)=\frac{<A(t)A(0)>}{<A(0)A(0)>},\]
from which we can describe most of the processes of interest, including
relaxation. We use the above conditions for the noise to obtain a self-consistent equation for $R(t)$ 
\begin{equation}
\label{self_consistent}
\frac{d R(t)}{d t}=-\int_0^t R(t-t')\Pi(t')\,d t'.
\end{equation}

If we then apply the Laplace transform (denoted by a tilde) to eq.~(\ref{self_consistent}), we get
\begin{equation}
\label{laplace_R}
\widetilde{R}(z)=\frac{1}{z+\widetilde{\Pi }(z)}.
\end{equation}
From the analysis of this equation, it is possible to obtain plenty of information concerning the asymptotic behaviour of the system. 
In order to make progress, we need to make some assumptions about
the origin of the memory. The direct way is to connect the random
force, $F(t)$, to a thermal bath composed of harmonic oscillators~\cite{Morgado02,Costa03,Bao03}. Therefore, for a system in contact with a heat reservoir (canonical), the memory becomes
\begin{equation}
\label{memory}
\Pi (t)=\int \rho (\omega )\cos (\omega t)\,d \omega ,
\end{equation}
where \( \rho \) is the noise density of states. The memory is clearly even for any noise distribution.  We shall use a coloured noise given by a generalization of the Debye spectrum\begin{equation}
\label{noise_dos}
\rho (\omega )=
\begin{cases}
\frac{2\gamma }{ \pi } \left( \frac{ \omega }{\omega_D } \right )^\beta &, \text{ if } \omega<\omega_D \\
0 &,\text{ otherwise},
\end{cases}
\end{equation}
with $\omega_D$ as a Debye cutoff frequency. The motivation for considering such cases arises from previous studies~\cite{Morgado02,Costa03} in which it was proved that if $\widetilde{\Pi}(z) \propto z^\nu$ as $z\rightarrow 0$, then the diffusion exponent is given by 
\begin{equation}
\label{diff_exponent}
\alpha=1+\nu.
\end{equation}

 \section{Memory, correlation functions and noise\label{sec.memory}} 
 In order to obtain $R(t)$, we state that its derivative must vanish at the origin, due to eq.~(\ref{self_consistent}); therefore, the correlation function can be neither an exponential nor a stretched exponential. The analytical continuation of the Laplace transform of an even function is an odd function, and vice-versa. From eq.~(\ref{laplace_R}), we see that $\widetilde{R}(z)$ is odd, because $\Pi(t)$ is even (see eq.~(\ref{memory})). Following the same line of thought, we arrive at the conclusion that $R(t)$ is even.
 Lee~\cite{Lee83} has also shown that both the memory and  the correlation function must be even functions for any Hamiltonian system. Consequently, we can write

\begin{equation}
\label{R2}
R(t)=\sum _{n=0}^{\infty }a_{n}t^{2n},
\end{equation}
with $ a_{0}=R(0)=1$ and,
\begin{equation}
\label{G2}
\Pi (t)=\sum _{n=0}^{\infty }b_{n}t^{2n}.
\end{equation}
We replace those, eqs.~(\ref{memory}) and (\ref{noise_dos}) in eq.~(\ref{self_consistent}) to obtain the following recurrence relation	
\begin{equation}
\label{an2}
a_{n}=-\frac{2\gamma \omega _{s}}{\pi (2n)!}\sum _{l=0}^{n-1}\frac{(-1)^{l}[2(n-1-l)]!\,\omega _{s}^{2l}}{(2l+1+\nu )}a_{n-1-l},
\end{equation}
 which displays a complex behaviour, where every order depends
on all previous ones. This is not a surprise for a non-Markovian system and  it is very useful to have it made explicit in a power series.

The power series defined by eq.~(\ref{R2}) is convergent: it is straightforward to prove by induction that $r_{n}=a_{n}/a_{n+1}<0$. Besides that, it is easy to show that the absolute values of $a_n$ decrease for $n>n_0$, for a certain $n_0$. Consequently, since eq.~(\ref{R2}) is an alternating series, and $r=\lim_{n\rightarrow \infty} |r_n| \rightarrow \infty$, it is possible to conclude that it converges~\cite{Apostol65} for every value of $ t$. 
By playing with the parameters $\nu$, $\omega_D$, and $ \gamma $, it is possible to generate a large number of functions.

\section{Normal diffusion\label{sec.normal_diffusion}}
Before we analyze anomalous diffusion, it is very useful to give a more
detailed description of normal diffusion. If one takes eq.~(\ref{noise_dos})
with $\beta =0$ and $\omega_D$ finite, diffusion is normal, despite the noise being coloured. Using this
noise in eq.~(\ref{memory}), we get $\Pi(t)=(2\gamma/\pi)\sin(w_s
t)/t$. Applying the Laplace transform yields $\widetilde{\Pi}(z)=(2\gamma/\pi) \arctan (\omega _{s}/z) $. It is interesting to note that this memory function is realized in electron gas models~\cite{Lee84}.	
We now show that it is possible to find at least three time scales for this memory. 
The first one is the short time scale, given by $\tau_D=\omega_D^{-1}$. 
In the short time scale, given by $t\ll \tau _{s} $, the first term in the expansion, eq.~(\ref{R2}), gives
\begin{equation}
\label{R3}
R(t) \sim \cos (\omega _{0}t),
\end{equation}
where $\omega _{0}=\sqrt{\Pi(0)}$. For this particular case, $\Pi(0)=2\gamma \omega _{s}/\pi$ introduces the intermediate time scale $\tau^{*}=\omega _{0}^{-1}\propto \sqrt{\tau_0 \tau_D}$. In the long time scale, $t\gg \tau_D$, or, equivalently, $z\ll\omega _{s}$, then $\widetilde{\Pi}(z)\sim \widetilde{\Pi}(0)=\gamma$. Therefore, from eq.~(\ref{laplace_R}), we obtain $R(t)\sim \exp(-t/\tau_{0})$, with $\tau_{0}=\gamma^{-1}$.
This result is the same as the one obtained from the normal Langevin equation, which can be obtained from the limiting case $\omega_D \rightarrow \infty$.

\begin{figure}
\centerline{\includegraphics[width=6.5cm,angle=270]{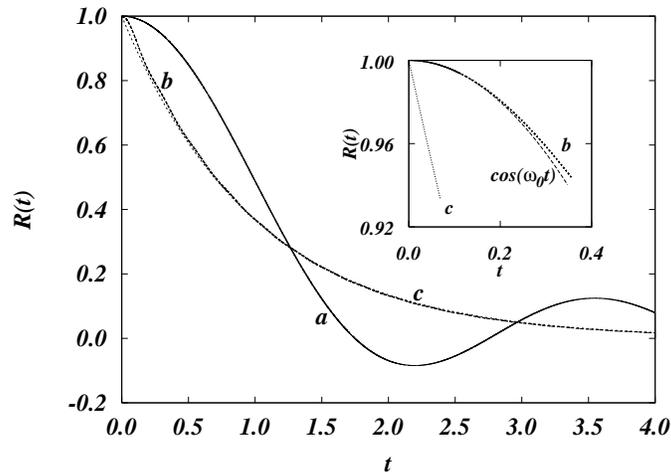}}
\caption{Plot of the correlation function $R(t)$ for normal diffusion
($\nu=0$). Here, $\gamma=1$ and $\omega_D=2 $ and $20$, for curves $a$ and
$b$, respectively. Curve $c$ is the plot of $\exp(-\gamma t)$. Note that
curves $b$ and $c$ are indistinguishable, except near the origin (see inset). Inset: Plot of curves $b$, $c$ and $\cos(\omega_0 t)$, near the origin.}
\label{fig1}
\end{figure}


In fig.~\ref{fig1} we plot the correlation function $ R(t)$ as a function of $t$ for normal diffusion. We use $\gamma =1$, and $\omega_D=2$ and $20$ in curves $(a)$, and $(b)$, respectively. Curve $(c)$ is the plot of the function $\exp(-\gamma t)$.  It is clear that curve $(a)$ is not an exponential; however, in many practical situations it is fine to approximate curve $(b)$ by an exponential for times larger than $\tau_D$. For short times, $t\ll \tau_D$, eq.~(\ref{R3}) is the solution and we disregard the exponential as the solution in both cases. The inset shows the correlation function $R(t)$ near the origin for $\omega_D=20$. Together with curve $(b)$ and $(c)$, we plot $\cos(\omega _0t)$. In short, ``good exponentials'', such as curve $(c)$ cannot exist as a solution in this case. In many experimental situations, the ratio $ \omega _{s}/\gamma$ can be very large. Nevertheless, we propose that there will always be a short time behaviour which permits distinction from an exponential.  Even if the distinction is not accessible to the experimental probe, it is very important to bear in mind that the correlation function is even.

\section{Anomalous diffusion}
\label{sec.anomalous_diffusion}

\begin{figure}
\centerline{\includegraphics[width=6.5cm,angle=270]{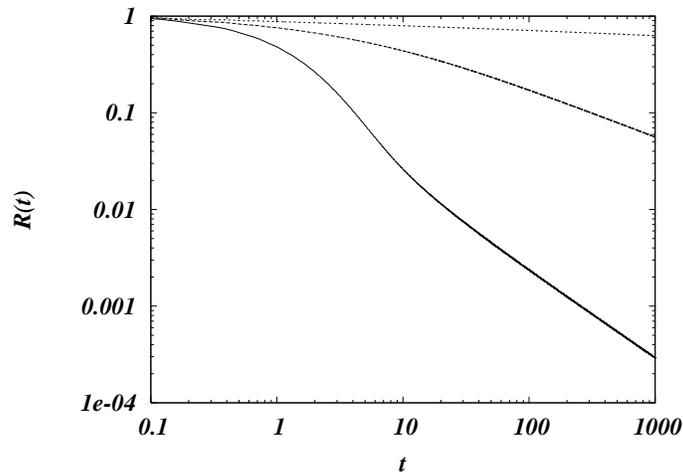}}
\caption{Log-log plot of the correlation function $R(t)$ for $\omega_D/\gamma=20$ and $\nu=0.9$, $0.5$, and $0.1$ from top to bottom. Notice the transition from a stretched exponential behaviour to a power law.}
\label{fig2}
\end{figure}

Now let us return to the general memory. By substituting eq.~(\ref{noise_dos}) into eq.~(\ref{memory}) with $\beta \neq 0$ and calculating the Laplace transform, we get

\begin{equation}
\label{G3}
\widetilde{\Pi }(z)=\frac{2\gamma }{\pi }\Psi _{\beta }(\omega _{s}/z).
\end{equation}
Here,
\begin{equation}
\label{psi}
\Psi _{\beta }(x)=x^{-\beta }\int _{0}^{x}\frac{y^{\beta }}{1+y^{2}}\,d y = \frac{1}{2}x\Phi(-x^2,1,\frac{1+\beta}{2}),
\end{equation}
where $\Phi(x,a,b)$ is the Lerch Transcendent~\cite{Erdelyi81}. This function has been found most useful in certain quantum statistics~\cite{Ciccariello04}.
This memory function has an asymptotic behaviour of the following form
\begin{equation}
\label{LimGa}
\lim _{z\rightarrow0 }\widetilde{\Pi }(z) \sim \tau _{\nu }^{\nu -1}z^{\nu },
\end{equation}
where
\begin{equation}
\label{nu}
\nu = 
\begin{cases}
\beta \text{,   for } \beta \leq 1 \\
1 \text{,   otherwise}.
\end{cases}
\end{equation}
This cutoff shows that the exponent $\alpha$, eq.~(\ref{diff_exponent}), cannot be larger than $2$
(ballistic motion). Recent results~\cite{Bao03} confirm that
cutoff; they obtain  $\nu=1$ when $\beta=2$.

By inverse transforming eq.~(\ref{laplace_R}), using eq.~(\ref{LimGa}), we get
\begin{equation}
\label{R1}
R(t)\sim E_{1-\nu }(-(t/\tau _{\nu })^{1-\nu }),
\end{equation}
which is a useful approximation for $t>\tau_D$. Here $E_{\delta}(x)$ is the Mittag-Leffler function~\cite{Mittag-Leffler05,Metzler00}.
This function behaves as a stretched exponential for short times
and as an inverse power law in the long time regime (see fig.~\ref{fig2}). Due to the asymptotic power-law behaviour 
\begin{equation}
E_{1-\nu }(-(t/\tau _{\nu })^{1-\nu })\sim [(t/\tau_\nu)^{1-\nu}\Gamma(-\nu)]^{-1},
\end{equation}
for $0<\nu<1$, we obtain an infinite diffusion constant $D$, which is a sign
of superdiffusion~\cite{Morgado02}. The power law exponent $1- \nu=2-
\alpha$ is in the interval $(0,2)$ as observed long ago by Friedrich Kohlrausch. For $\nu =0$, the Mittag-Leffler function is the usual exponential and $\tau _0$ is the relaxation time. The limiting cases correspond to $\nu=1$ and $\nu=-1$; for these values, we have $E_0(x)=(1-x)^{-1}$ and $E_2(x)=\cosh(\sqrt{x})$, respectively. This last case corresponds to the limit of subdiffusion, \emph{i.e.}, $\alpha=0$. Since the argument of the function is negative, the Mittag-Leffler function is in this case a cosine.
For short times, $t \ll \tau_D$ the first term in the expansion gives a result identical to eq.~(\ref{R3}), with $\omega_\nu^2=2\gamma \omega_D/(1+\nu)\pi$.

In fig.~\ref{fig2}, we display the correlation function $R(t)$ as a function
of t for anomalous diffusion, obtained from the numerical integration of
eq.~(\ref{self_consistent}). We select broadband noise with $\omega_D/\gamma=20$. From top to bottom,  $\nu=0.9, 0.5$ and $0.1$. We notice the transition from stretched exponential to power law behaviour. Note as well that as the behaviour becomes more anomalous, \emph{i.e.} as $\nu$ increases, the relaxation becomes slower.   

\section{Markovian vs. non-Markovian behaviour}
\label{sec.markovian}

Recently some experimental situations have been found where non-Markovian processes display Markovian behaviour after a long time. In particular, we call attention to the experiments of Merikoski \emph{et al.} about flame evolution~\cite{Merikoski03}. We show now that this transition is possible for normal diffusion. For large $t$, eq.~(\ref{self_consistent}) can be written as
\begin{equation}
\label{approx_corr}
\frac{d R(t)}{d t}\approx -R(t)\int_0^t \Pi(t')\,d t'.
\end{equation}
We can justify this approximation by the following arguments: if $t'\ll t$,
then $R(t-t')\approx R(t) $ and for those values $\Pi(t')$ gives the major contribution. As $t'$ increases, say $t'\approx t/2$, both $R$ and $\Pi$ are small. Finally, for $t'\approx t$, $R(t-t')\approx 1$, but $\Pi(t')$ is very small. 

Notice that this decoupling happens in many situations in physics, otherwise
all correlated systems would present aging. For normal diffusion, the convergence of the integral of the memory function to a constant value $\widetilde{\Pi}(z\rightarrow 0)=\gamma$ makes the decoupling acceptable. In that case, for broadband noise $R(t)\sim  \exp(-\gamma t)$,
as expected (see fig.~\ref{fig1}). Consequently, as long as the diffusion is normal, $\gamma=\widetilde{\Pi}(0)$ is finite independently of the character of $\Pi(t)$; after a time $t>\gamma^{-1}$ all the processes behave as those governed by the usual Langevin equation. For anomalous diffusion, the dependence on time of the integral of the memory in  eq.~(\ref{approx_corr}) makes the decoupling fail. In short, the exponential behavior in normal diffusion is associated with time invariance while the non-exponential behaviour is associated with an aging process which is in essence non-Markovian.

\section{Conclusion}
\label{sec.conclusion}
We discussed the relaxation properties of a dynamical variable $A(t)$ and we proved that neither exponentials nor stretched exponentials can give a full description of the relaxation process. However, these functions can be the asymptotic behaviour found in many experimental and theoretical works and we found an even function that can approach an exponential as we broaden the noise (increase the ratio $\omega_D/\gamma$). For broad noise, the correlation function for anomalous diffusion will be a stretched exponential and finally a power-law.

Fractional derivatives have been successfully applied to the study
of subdiffusive behaviour~\cite{Metzler00}. Those yield $\alpha =1+\nu $,
for $-1<\nu <0$. The GLE formalism allows $\nu$  to assume the values in the range~\cite{Morgado02,Costa03,Pottier04} $ -1<\nu <1$, which encompass both subdiffusion and superdiffusion. There is a great advantage in our formulation, since the noise density of states exists not only in systems governed by a GLE, but in most physical systems. Also, we avoid the levy function which has infinite mean square deviation.

As we have already mentioned, glassy systems seem to be a rich field for studying these phenomena. It would be very helpful if the relaxation process could be studied and exponent $\alpha $ determined for those diffusive processes.
Other related phenomena are chaos synchronization~\cite{Longa96,Ciesla01} and anomalous
reaction rates\cite{Oliveira95,Oliveira98a,Oliveira98,Maroja01}, which we expect to discuss soon. 

\acknowledgments
This work was supported by CAPES, CNPq and FINATEC.

\bibliographystyle{unsrt}

\end{document}